\documentclass[conference]{IEEEtran}
\IEEEoverridecommandlockouts
\usepackage{cite}
\usepackage{amsmath,amssymb,amsfonts}
\usepackage{algorithmic}
\usepackage{graphicx}
\usepackage{textcomp}
\usepackage{xcolor}
\usepackage{filecontents,catchfile} 
\def\BibTeX{{\rm B\kern-.05em{\sc i\kern-.025em b}\kern-.08em
    T\kern-.1667em\lower.7ex\hbox{E}\kern-.125emX}}
\begin{document}

\title{Design Optimization of a Three-Phase Transformer Using Finite Element Analysis\\
\thanks{All authors contributed equally to this work.}
}

\author{\IEEEauthorblockN{Ahmet Furkan Hacan\IEEEauthorrefmark{1}, Bilal Kabas\IEEEauthorrefmark{1}, Samet Oguten\IEEEauthorrefmark{1}}
\IEEEauthorblockA{\IEEEauthorrefmark{1}Department of Electrical-Electronics Engineering,
Abdullah Gül University\\
Kayseri, Turkey\\}
Email: ahmetfurkan.hacan@agu.edu.tr, bilal.kabas@agu.edu.tr,
samet.oguten@agu.edu.tr
}

\maketitle

\begin{abstract}
Optimization of design parameters of a transformer is a crucial task to increase efficiency and lower the material cost. This research presents an approach to model a three-phase transformer and optimize design parameters to minimize the volume and loss. ANSYS Maxwell 2D is used to model the transformer and analyze it for different design parameters. The multi-objective differential evolution algorithm is used to find optimum design parameters that minimize the volume and loss. In this paper, we present the optimum design parameters for a 1 kVA transformer with a particular input and output voltage specification. The transformer with these optimum design parameters is then tested for different loading conditions and power factor values. The results show that the maximum efficiency is obtained for 75\% loading condition with unity power factor. As the power factor decreases, the efficiency decreases as well.
\end{abstract}

\section{Introduction}

Transformers are crucial components of power electronics. Many application requires different voltage levels. A step-up transformer is used to increase AC voltages while a step-down transformer is used to decrease. Transformers have a vital role in power transmission and distribution systems as transmission lines require high voltages to mitigate copper losses. On the other hand, these high voltage levels should be reduced to be used at homes for safety. The aim of this research is to optimize the design parameters of a three-phase transformer through 2D modeling in ANSYS Maxwell. As Maxwell 2D is used for analysis to obtain efficiencies, the multi-objective differential evolution algorithm is used to optimize design parameters to minimize the volume and loss.

The design of three-phase transformers is usually made with three legs. The primary and secondary winding are wound on these transformer legs. In transformers, high voltage is wound over low voltage winding. There is no direct link between the secondary winding and primary winding. The connection between primary and secondary winding is provided by magnetic effects on the core. Winding, where the input voltage is supplied, creates a magnetic flux on the core. The magnetic flux on the core provides current and voltage on the other winding. In this way, the power is transferred from one coil to another. The finite element method (FEM) is used to analyze the magnetic effects on the transformer. The logic of the finite element method (FEM) is to divide the model into numbered elements, analyze each element, and then combine these results on the mesh.
\section{related work}
In literature, there are many studies on transformers since transformers are considered to be the heart of the transmission and distribution system \cite{Georgilakis2009}. Therefore, it is highlighted that design of high-quality transformers is crucial for obtaining low-loss, low-cost systems. Moreover, in a study, it is suggested that almost 7.4\% of the power produced in the world is lost in transmission and distribution systems \cite{Alhan2017}. Thus, it is important to make these systems as efficient as possible. On the other hand, increasing efficiency often results in increased weight and size leading to more expensive production. The materials used in a transformer are quite expensive that they constitute a large portion of the cost of the transmission and distribution systems. Therefore, it is needed to find the optimum design that both has high efficiency and low size or weight. 

Considering the aforementioned reasons, the design optimization of a transformer takes the attention of engineers. However, this optimization is not a straightforward task. The objective of designing a highly efficient, low-cost transformer involves complex, non-linear functions \cite{Alhan2017}. Therefore, it is highlighted that analytical tools may not be accurate enough and numerical methods such as the finite element method (FEM) should be used to design and analyze transformers \cite{akpojedjeparametric}. 

In a study \cite{akpojedjeparametric}, for the optimization task of three-phase transformer, an optimization algorithm is used alongside FEM to find the optimum design and overcome the problem of complexity of designing transformers. The algorithm used in this study is the Genetic Algorithm (GA) which relies on the ‘biological evolution theory of survival of the fittest. In another study, it is stated that ‘differential evolution’ which is a similar optimization algorithm to GA, converges a faster and more robust algorithm which is successfully applied to power systems \cite{Georgilakis2009}.

\section{Method}
\begin{figure}[t]
    \centerline{\includegraphics[width=.9\linewidth]{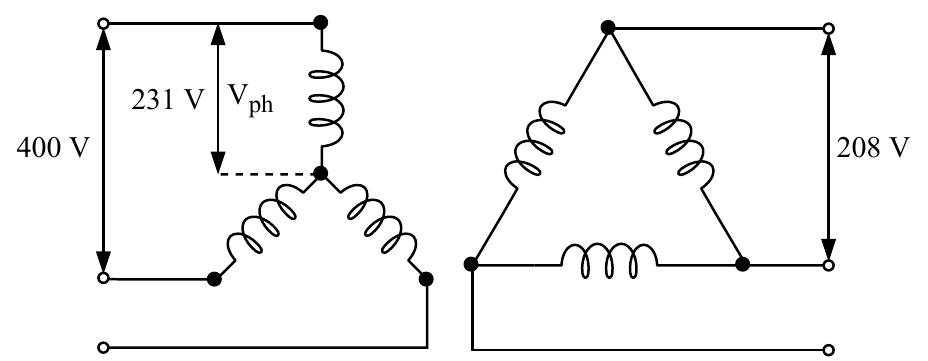}}
    \caption{Winding connections, line, and phase voltages of the transformer.}
    \label{trans_conf}
\end{figure}

As per design requirements, the transformer should have 1 KVA power rating and operate at 50 Hz. For a 400 V input voltage, it should output 208 V. This transformer is a step-down transformer. In the design, Y-$\Delta$ configuration is used as shown in Fig.~\ref{trans_conf}. Therefore, the phase voltage at the primary side is 231 V.

\subsection{2D Modeling}

\begin{table}
\caption{The Optimized Transformer Design Parameters for Maximum Efficiency and Minimum Volume}
    \begin{center}
        \begin{tabular}{|c|c|c|c|}
        \hline
        & \multicolumn{3}{|c|}{\bf Copper Properties} \\
        \cline{2-4}
            {\bf Coppers} & {\bf Max Current} & {\bf Area$^{\mathrm{a}}$} & {\bf Resistance} \\
        
                   &        \bf (A) &      \bf (mm\textsuperscript{2}) & \bf (ohms/km) \\
        \hline
        {\bf Coil 1 Copper} &        1.5 &      0.518 &      26.41 \\
        \hline
        {\bf Coil 2 Copper} &        1.8 &      0.653 &      33.29 \\
        \hline
        \multicolumn{4}{l}{$^{\mathrm{a}}$Cross sectional area of the copper.}
        \end{tabular}  
    \end{center}
    \label{copper_properties}
\end{table}

\begin{figure}[t]
    \centering
    \includegraphics[width=.6\linewidth]{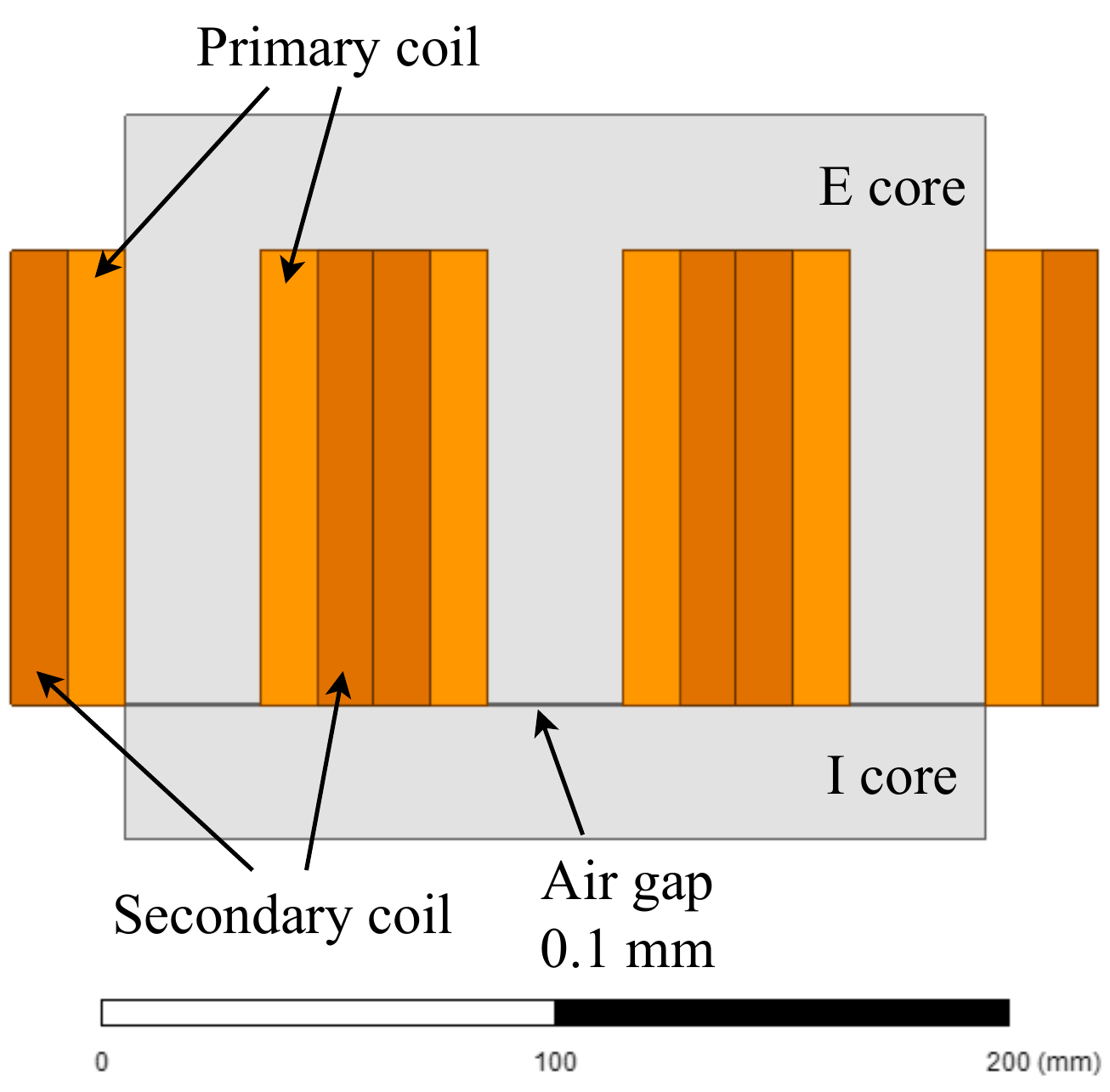}
    \caption{The design of the three-phase transformer in Maxwell 2D.}
    \label{trans_model}
\end{figure}

\begin{figure}[t]
    \centering
    \includegraphics[width=.6\linewidth]{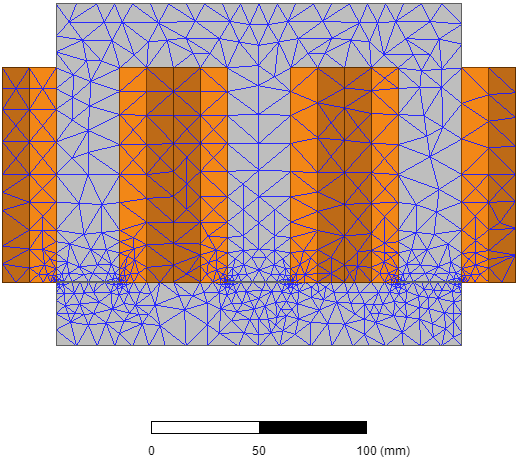}
    \caption{Mesh of the transformer model in Maxwell 2D.}
    \label{trans_mesh}
\end{figure}

\begin{figure}[t]
    \centerline{\includegraphics[width=.6\linewidth]{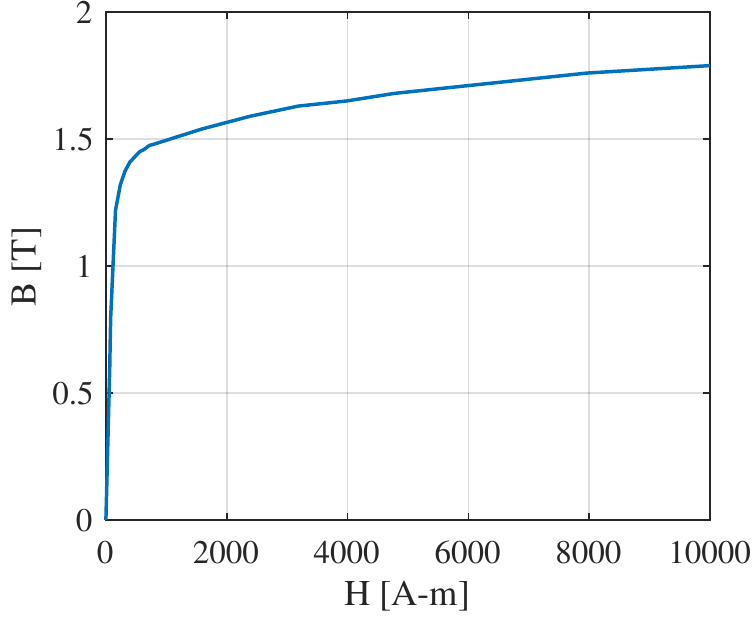}}
    \caption{B-H curve of the M15 grade steel forming the transformer core.}
    \label{BH_curve}
\end{figure}

The transformer is modeled in Maxwell 2D as shown in Fig.~\ref{trans_model} and the mesh of the model is shown in Fig.~\ref{trans_mesh}. In the design, the M15 grade steel is used as the core material. The B-H curve of this material is shown in Fig.~\ref{BH_curve}. The magnetic field is saturated for values exceeding approximately 1.8 Tesla. The core consists of E and I parts. There is a 0.1 mm air gap between these two parts. There are three primary coils and three secondary coils. The secondary coils are wound around the primary coils. According to our initial experiments, using all the window area for coils result in less material cost. Therefore, in the model, there is no space between the coils and core. It should be noted that in the 2D model, this does not cause a conduction path problem. It is assumed that half of the window area is filled with the primary coil and the other half is filled with the secondary coil. In fact, the turns ratio is 1.11, and copper sizes are different for the primary and secondary coils therefore the actual areas that primary and secondary coils take are not equal. However, this difference has an insignificant effect on the results. The thickness of the core is constant everywhere. Since an optimization algorithm is used to find optimum design parameters, core dimensions which are height, width, thickness, and depth values are parameterized in Maxwell 2D. In addition to these, the voltage regulation $(V_{R})$ value is also parameterized. The voltage regulation parameter is used to regulate the excitation voltage thus, the phase voltage at the primary winding becomes approximately 231 V and hence the peak excitation voltage is calculated using \eqref{peak_excitation_volt}.
\begin{equation}
    V_{peak} = V_{1,phase} \times \sqrt{2} + V_{R} \quad (V)
    \label{peak_excitation_volt}
\end{equation}
To ensure that the input power is 1 KVA, the load resistance should be correctly found. Considering a unity power factor and 100\% loading condition, for 1 KVA power rating with 231 V primary voltage and 208 V secondary voltage, primary and secondary side currents can be estimated as 1.44 A and 1.6 A. Therefore, the total resistance on the secondary side including winding and load resistances should be 130 $\Omega$. It is also important to determine winding resistances to obtain more realistic analysis results. Winding resistances depend on the copper cross-sectional area and length of the copper and hence they depend on core dimensions, specifically window area, number of turns, and the maximum value of the current flow. The window area is calculated using \eqref{Aw_calc} where $W_c$, $H_c$, and $T_c$ are width, height, and thickness of the core, respectively. For the estimated maximum current values of the primary and secondary winding, 1.44 A and 1.6 A, respectively, appropriate copper cross-sectional area and resistance values are given in Table \ref{copper_properties}. Coil turns depend on the window area and it is assumed that copper can fill 35\% of a given area. Therefore, coil turns, $N_1$ and $N_2$, are calculated using \eqref{turn_calc} where $A_{cop_1}$ and $A_{cop_2}$ are copper cross sectional areas given in Table \ref{copper_properties}.
\begin{equation}
    A_{w} = \frac{W_{c}-3T_{c}}{2} \times (H_{c}-2T_{c}) \quad (mm^{2})
    \label{Aw_calc}
\end{equation}
\begin{equation}
    \frac{A_{w}}{2} \times 35\% = N_{1} \times A_{cop_{1}} + N_{2} \times A_{cop_{2}}
    \label{turn_calc}
\end{equation}
Considering copper properties given in Table \ref{copper_properties}, primary and secondary winding resistance can be calculated using \eqref{R_pri_calc} and \eqref{R_sec_calc} where $D_c$, $T_c$, and $W_w$ are core depth, core thickness, and window width.
\begin{equation}
    R_{pri} = (D_{c} + T_{c} + \frac{W_{w}}{2}) \times 2N_{1} \times 33.29 \times 10^{-3} \quad (\Omega)
    \label{R_pri_calc}
\end{equation}
\begin{equation}
    R_{sec} = (D_{c} + T_{c} + \frac{3W_{w}}{2}) \times 2N_{2} \times 26.41 \times 10^{-3} \quad (\Omega)
    \label{R_sec_calc}
\end{equation}
%

Instead of building an external circuit to introduce a load, the secondary winding resistance is calculated using \eqref{R_tot_calc} so that it also includes the load resistance.
\begin{equation}
    R_{tot} = R_{res} + R_{load} \quad (\Omega)
    \label{R_tot_calc}
\end{equation}
To obtain different power factor values such as 0.9 and 0.8, the load inductance is calculated using \eqref{load_inductance} where Z is the total impedance calculated using input power and winding voltages and its value is 130 $\Omega$. Winding inductance does not present in this expression as winding inductance are all neglected in this research.
\begin{equation}
    L_{load} = \frac{ \sqrt{Z^{2}-R_{tot}^{2}} }{2\pi f} \quad (H)
    \label{load_inductance}
\end{equation}
\subsection{Optimization of design parameters}
Multi-objective Differential Evolution Algorithm (MODE) is chosen as the optimization method that will be used with FEA \cite{Storn1997}. As it is described in the preceding sections, evolutionary algorithms mimic the biological evolutionary theory of survival of the fittest. The process of MODE begins with generating the initial candidate design followed by the analysis of candidate design and generating a second set of candidate designs called ‘child vector’. As the child vector is generated and analyzed on ANSYS Maxwell, the first set of candidate designs and child vectors are compared to pick the ones that should survive in the next generation. The survived designs are selected considering the costs and losses. This process is repeated a number of times and it is aimed for designs to converge high efficiency and small size points. The design parameters of the model that are aimed to be optimized are height, width, depth, thickness of the limbs, and voltage regulation. The parameter boundaries are determined through a heuristic approach. We used MATLAB to run the differential evolution algorithm.

\subsection{Efficiency and Cost Analysis}
Efficiency in transformers is the ratio of the power in the secondary winding and the power in the primary winding. In order to have high efficiency in transformers, copper loss and core loss should be minimized. In this research, two methods are followed for efficiency calculation. The reason is that efficiency calculation is required for different power factor values. One of the methods used for efficiency calculation is done for the power factor value 1, while the other method is used for the power factor value less than 1. When the power factor value is 1, the inductance value of the load is assumed to be zero in the efficiency calculation. The output power required for efficiency calculation is found from the power released on the load. For its input power, it is found by multiplying the phase voltage of the transformer with the current. Then the output power is divided into input power to find the efficiency. 

In this research, inductance is added to the circuit so that the power factor is less than 1 and lagging. In this case, the efficiency calculation is performed according to the other case. The calculation of input power is found by multiplying phase voltage, current, and power factor. The only difference from the method when the power factor is equal to 1 is to multiply the resulting value by the power factor. The reason is that apparent power emerges when it is not multiplied by the power factor. The output power is the power consumed on the load. But the impedance value of the added coil should be reduced from the resistance value of the load. The reason is to keep the total impedance value constant on the secondary winding.

\begin{figure*}[t]
    \centering
    {\includegraphics[width=0.24\linewidth]{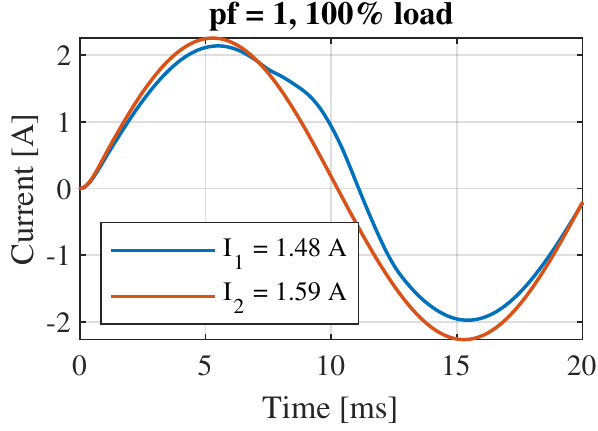}}
    \hfill
    {\includegraphics[width=0.24\linewidth]{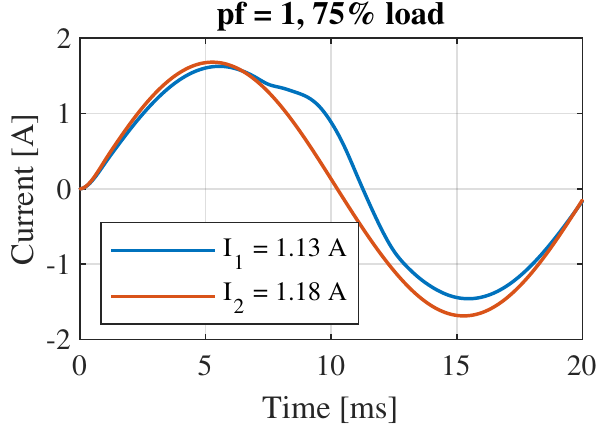}}
    \hfill
    {\includegraphics[width=0.24\linewidth]{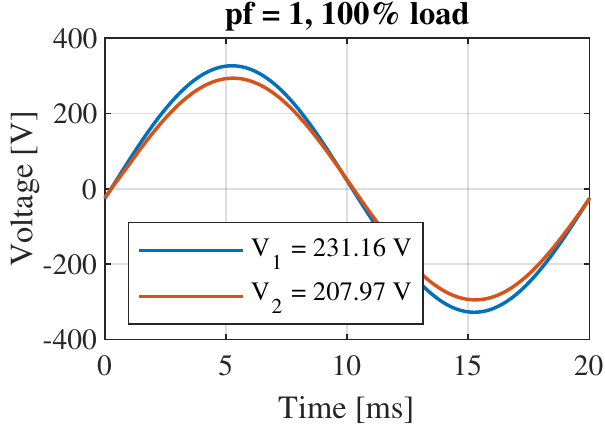}}
    \hfill
    {\includegraphics[width=0.24\linewidth]{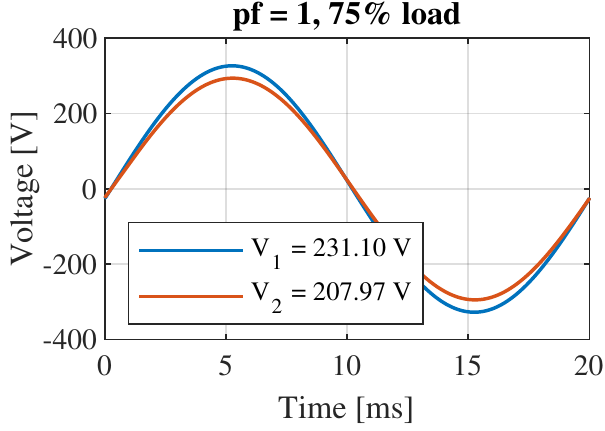}}
    \par\medskip
    
    {\includegraphics[width=0.24\linewidth]{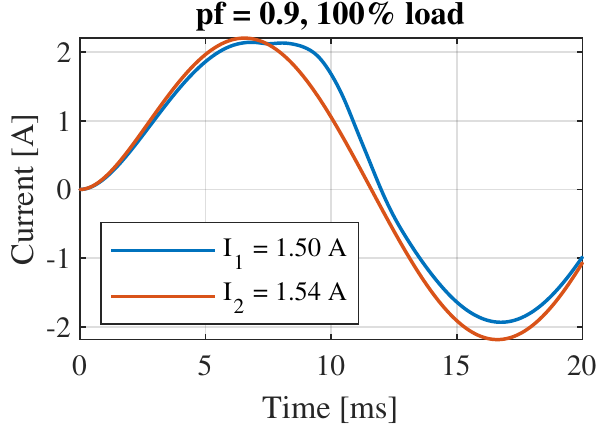}}
    \hfill
    {\includegraphics[width=0.24\linewidth]{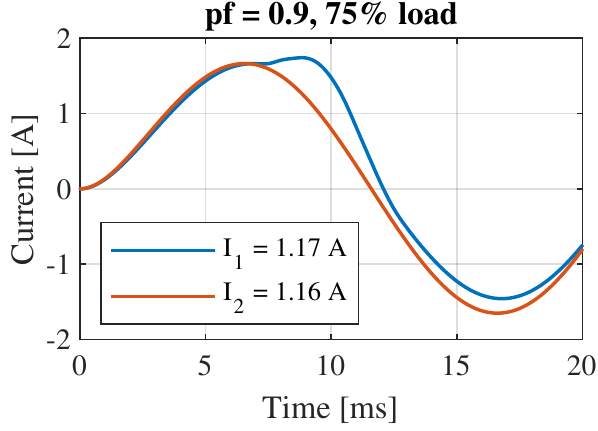}}
    \hfill
    {\includegraphics[width=0.24\linewidth]{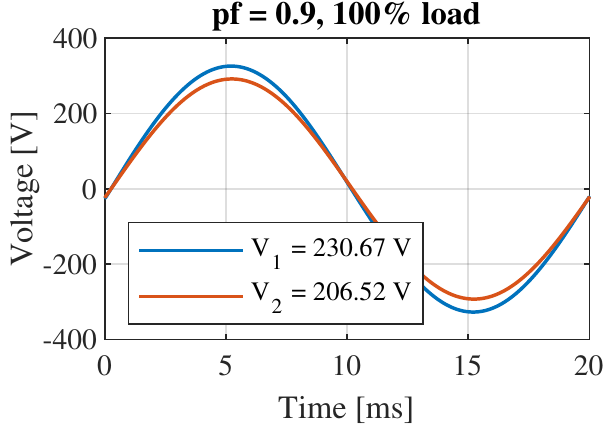}}
    \hfill
    {\includegraphics[width=0.24\linewidth]{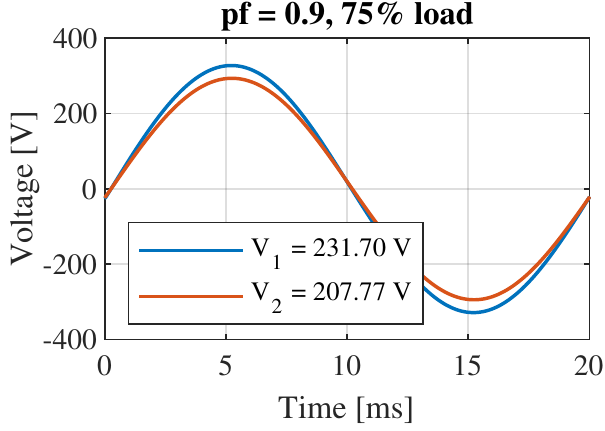}}
    \par\medskip
    
    {\includegraphics[width=0.24\linewidth]{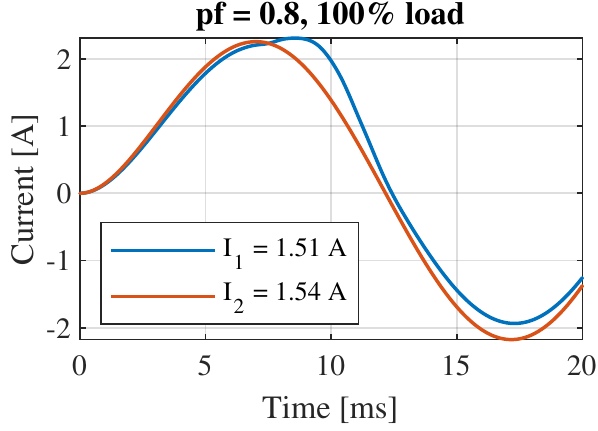}}
    \hfill
    {\includegraphics[width=0.24\linewidth]{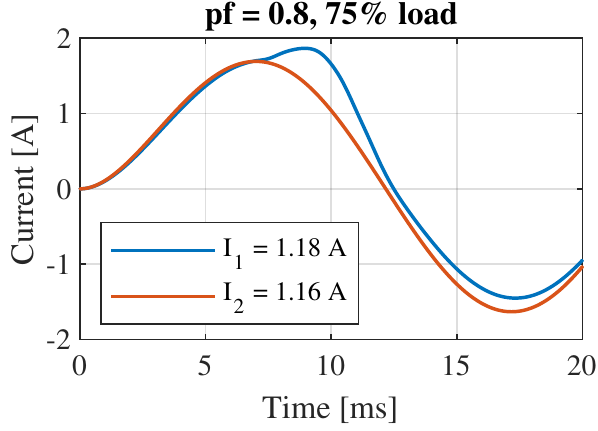}}
    \hfill
    {\includegraphics[width=0.24\linewidth]{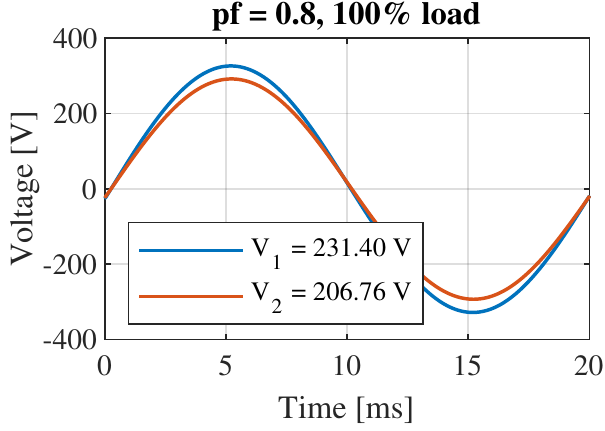}}
    \hfill
    {\includegraphics[width=0.24\linewidth]{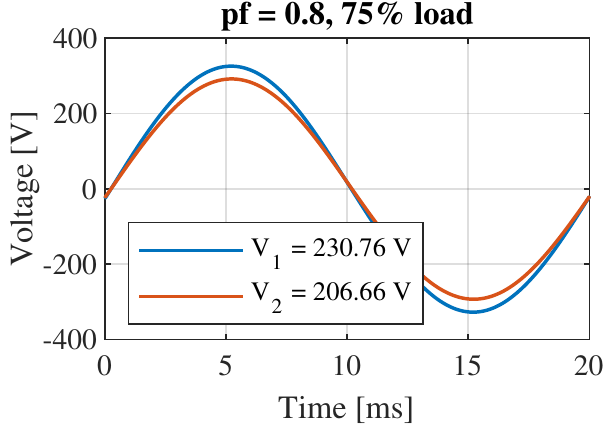}}
    \par\medskip
\caption{Induced voltages and currents at primary and secondary winding for power factor values of 1, 0.9 and 0.8 and for 100\% and 75\% loading conditions.}
\label{volt_curr_graphs}
\end{figure*}

In addition to high efficiency, low cost is aimed. The cost is proportional to the volume of the transformer. Transformer volume is calculated with copper volume and core volume. In order to calculate copper volume, the cross-sectional area of cables determined according to the amount of current, thickness, depth, winding number, and legs of the core were used. In calculating the volume of the core, the cross-sectional area of the core and depth are used.

\subsection{Finding Equivalent Circuit Parameters}
The equivalent circuit parameters of the transformer were found with open circuit and short circuit tests. Theoretically, with the open circuit test, the resistance and inductance values of the core are found. The reason is that the load side of the transformer does not draw current in the open circuit test, so the current flows only through the core. With open circuit  test, we find phase currents. Since there is a difference in phase currents, we computed the contribution of each phase to the core loss considering the phase currents. Using \eqref{R_Core}, the resistance value of the core is calculated. Then the current value flowing through the core resistor is found. \eqref{Inductance_current} is used to find the value of current flowing through the inductor. 
\begin{equation}
    P_{core} = V_{1}^2/R_{c} \quad (W)
    \label{R_Core}
\end{equation}
\begin{equation}
    I_{L} = \sqrt{I^{2}-I_{R}^{2}} \quad (A)
    \label{Inductance_current}
\end{equation}
Using current and voltage values, the inductance is found by using \eqref{inductance}. In order to understand whether the parameter values found by this method are correct, firstly, the equivalent impedance is found from the resistance and inductance values of the core. Afterward, the required equivalent impedance value is calculated by means of phase voltage and current. There is no difference between the equivalent impedance value and the equivalent impedance value found according to the resistance and inductance value of the core. 

Using the short circuit test, the winding resistance and inductance values can be calculated since the equivalent impedance value of the core is significantly higher than the equivalent impedance value of winding. The current through the windings, and the resistance and inductance values of the windings are calculated using the short circuit test. In this research, the winding resistance value is obtained from the losses on windings by using \eqref{losses_power}. Then, using the current and resistance value, the voltage on the winding resistor is found. The value of the voltage on the inductance is found by the square root of the difference of the squares of the voltage values on the phase voltage and winding resistors. \eqref{inductance} is used to find the reactance value. In order to understand the accuracy of the parameter values found with the short circuit test, the equivalent impedance value is found using the resistance and inductance values. Then, when the equivalent impedance value is multiplied with the current value, the voltage value given as input is found.
\begin{equation}
    X_{L} = V_{L}/I_{L} \quad (\Omega)
    \label{inductance}
\end{equation}
\begin{equation}
    P_{loss} = I^2 \times R \quad (W)
    \label{losses_power}
\end{equation}
\section{Results}
\begin{table*}[htbp]
\caption{Analysis Results of the Transformer for Different Loading Conditions and Power Factor Values}
    \begin{center}
        \begin{tabular}{|c|c|c|c|c|c|c|c|c|c|c|c|c|}
            \hline
            &               \multicolumn{ 4}{|c|}{{\bf pf = 1}} &             \multicolumn{ 4}{|c|}{{\bf pf = 0.9}} &             \multicolumn{ 4}{|c|}{{\bf pf = 0.8}} \\
            \hline
            {\it {\bf Load (\%)}} & {\it {\bf 100}} & {\it {\bf 75}} & {\it {\bf 50}} & {\it {\bf 25}} & {\it {\bf 100}} & {\it {\bf 75}} & {\it {\bf 50}} & {\it {\bf 25}} & {\it {\bf 100}} & {\it {\bf 75}} & {\it {\bf 50}} & {\it {\bf 25}} \\
            \hline
            {\bf Efficieny (\%)} &      90.31 &      91.50 &      91.41 &      86.76 &      82.74 &      82.97 &      81.61 &      73.48 &      72.24 &      81.09 &      79.51 &      70.63 \\
            \hline
            {\bf V\textsubscript{1} (V)} &     231.16 &     231.10 &     231.70 &     230.85 &     230.67 &     231.70 &     231.38 &     231.06 &     231.40 &     230.76 &     230.28 &     231.20 \\
            \hline
            {\bf V\textsubscript{2} (V)} &     207.97 &     207.97 &     208.60 &     207.87 &     206.52 &     207.77 &     208.79 &     207.81 &     206.76 &     206.66 &     206.63 &     207.85 \\
            \hline
            {\bf I\textsubscript{1} (A)} &       1.49 &       1.13 &       0.79 &       0.47 &       1.51 &       1.18 &       0.84 &       0.52 &       1.51 &       1.19 &       0.85 &       0.53 \\
            \hline
            {\bf I\textsubscript{2} (A)} &       1.60 &       1.19 &       0.79 &       0.40 &       1.55 &       1.17 &       0.78 &       0.39 &       1.54 &       1.17 &       0.78 &       0.39 \\
            \hline
            {\bf V\textsubscript{R} (V)} &      12.00 &       9.00 &       7.00 &       3.00 &      10.00 &       9.00 &       6.00 &       3.00 &      10.00 &       7.00 &       4.00 &       3.00 \\
            \hline
            {\bf R (ohm)} &     130.00 &     174.75 &     263.25 &     521.22 &     117.00 &     156.00 &     234.00 &     468.00 &     104.00 &     138.67 &     208.00 &     416.00 \\
            \hline
            {\bf L (mH)} &       0.00 &       0.00 &       0.00 &       0.00 &     180.37 &     240.49 &     360.75 &     721.49 &     248.48 &     331.04 &     496.56 &     993.13 \\
            \hline
            {\bf P\textsubscript{in} (W)} &    1046.39 &     778.21 &     526.46 &     284.87 &    1016.32 &     779.62 &     535.45 &     301.24 &    1032.02 &     776.04 &     535.18 &     309.28 \\
            \hline
            {\bf P\textsubscript{out} (W)} &     944.99 &     712.06 &     481.25 &     244.31 &     840.99 &     646.87 &     436.98 &     221.36 &     826.98 &     629.29 &     425.55 &     218.43 \\
            \hline
        \end{tabular}  
    \end{center}
    \label{tab1}
\end{table*}

\begin{figure}
    \centerline{\includegraphics[width=.9\linewidth]{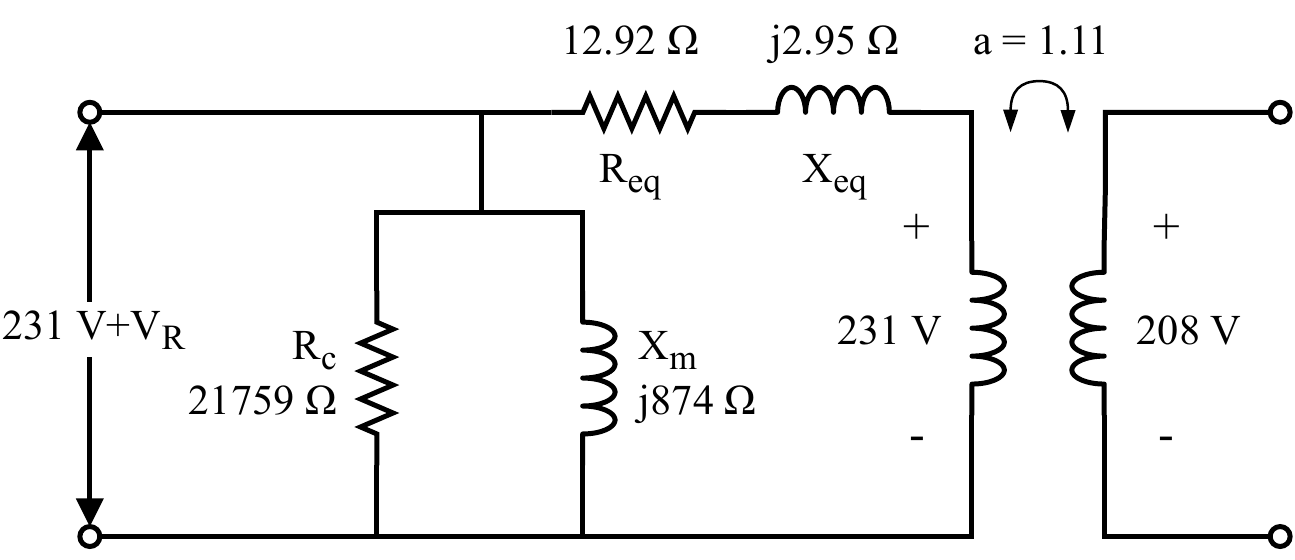}}
    \caption{Equivalent circuit parameters of the optimized transformer.}
    \label{eq_circuit}
\end{figure}

\begin{figure}
    \centerline{\includegraphics[width=.9\linewidth]{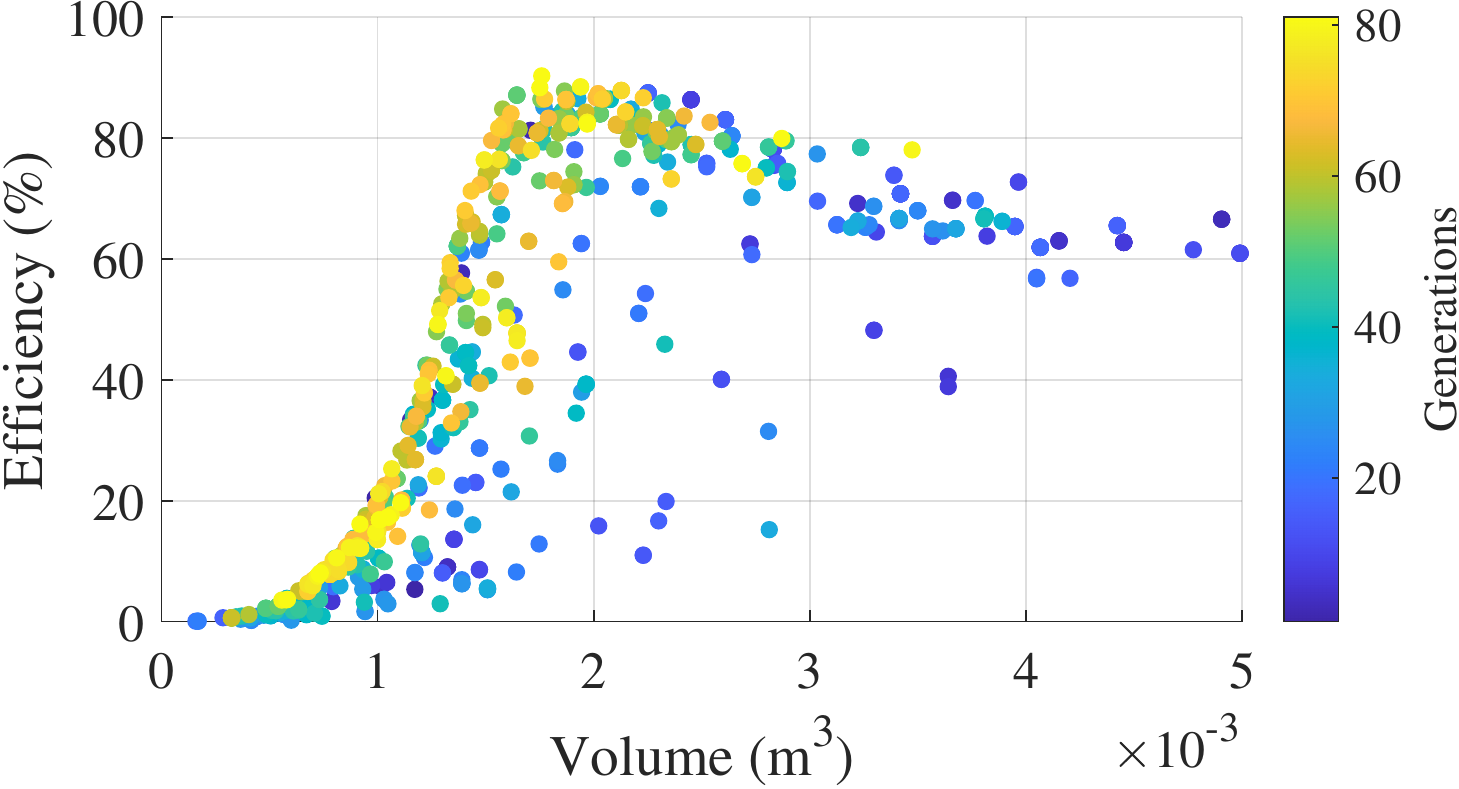}}
    \caption{Efficiency versus volume results of the differential evolution algorithm for 80 generations.}
    \label{de_results}
\end{figure}

The optimization is run for 80 generations, each having a population size of 12. Therefore, a total of 960 candidate designs are evaluated. The result of the optimization algorithm is shown in Fig.~\ref{de_results}. Furthermore, Fig.~\ref{de_results} shows how the candidate designs are evolved throughout the generations. Among these candidate designs, Table~\ref{optParams}, one of the optimum designs with high efficiency and low core and copper size, was selected. The chosen design has 90.31\% efficiency at rated load and $ 1.759 \times 10^{-3} \ m^{3}$ total volume of copper and core. The equivalent circuit parameters per phase have been found by applying the full load test and open-circuit test as shown in Fig.~\ref{eq_circuit}.

In Fig.~\ref{volt_curr_graphs}, current and voltage graphs of primary and secondary windings are shown according to the various power factor and load values. As seen in the graphs, as the load value decreases, the voltage value remains constant, but the current value has decreased. At the same time, when the power factor value and load value decrease, it has been observed that there are distortions in the sinusoidal form of the current on the side of the primary winding. Also, when the power factor value decreases, it has been observed that there is lagging between voltage and current.

Various analyses are shown according to different load and power factor values in Table \ref{tab1}. As seen in the table, as the percentage value of the load decreases, the resistance value increases. As a result, it is observed that the current values decreases on both the primary winding side and the secondary winding side. It is seen in Table \ref{tab1} that there is a decrease in both input power and output power because the voltage value remains constant while the current value decreases. As a result, when the power factor value is equal to 1 and the percentage value of the load is 75\%, the highest efficiency value is 91.50\%. The reason why the transformer reaches the highest efficiency value at 75\% load is that the loss on windings decreases as the current value decreases in windings.

\begin{table}[b]
\caption{The Optimized Transformer Design Parameters for Maximum Efficiency and Minimum Volume for Unity Power Factor and 100\% Loading Condition}
    \begin{center}
        \begin{tabular}{|c|c|c|c|c|}
        \hline
        \multicolumn{5}{|c|}{{\bf Optimum Design Parameters}} \\
        \hline
        \textbf{Height} & \textbf{Width} & \textbf{Depth} & \textbf{Thickness} & \textbf{V\textsubscript{R}} \\
        \bf (mm) & \bf (mm) & \bf (mm) & \bf (mm) & \bf (V) \\
        \hline
            159.76 &     189.64 &      52.12 &      29.88 &         12 \\
        \hline
        \end{tabular}  
         \label{optParams}
    \end{center}
\end{table}

\section{discussion}
One of the limitations of this research is that the core material has not been laminated. However, results suggest that Eddy current losses contribute a small portion of the total loss. One of the difficulties encountered in this research is that due to core and copper losses, the primary winding voltage differs significantly from the input voltage. To address this issue, a voltage regulation parameter is used so that the induced voltage at the primary winding can be kept constant. Yet another difficulty is that the multi-objective differential evolution algorithm used in optimization has tended to decrease the volume other than the total loss. However, this issue may be addressed by applying a better control structure for the selection of candidate designs.

To get better results in the research, some changes could be made in both the transformer model and the optimization algorithm. In the transformer model, thickness is considered equal everywhere. Defining more parameters to manipulate the thickness of the transformer for different regions of the core would allow the optimization algorithm to give better results. The optimization algorithm sends parameter values to ANSYS Maxwell to solve the transformer model. The range of values is determined for each of the parameters sent to ANSYS Maxwell. Adjusting this range of values to give the transformer model higher efficiency and lower cost could improve the model. Finally, the optimization algorithm needs to be run repeatedly to produce good results. For this reason, running the algorithm on a high-performing computer would help to converge to higher efficiency and lower volume values.
\section{Conclusion}
In this study, it is aimed to design and optimize a 1 KVA three-phase transformer. However, transformer design is not a straightforward process for engineers because of the complexities of calculations. Therefore, the finite element method was used which is widely used in many engineering applications including magnetic analysis. Concretely, an initial model is created on ANSYS Maxwell followed with parametrization of the model for optimization. With the help of MODE, the initial model is improved in terms of efficiency and manufacturing cost by minimizing the size of the transformer. The optimization algorithm was run for 80 generations that each of them had 12 candidate designs. Among the simulated candidate designs, one of them was selected having 90.31\% efficiency and $ 1.759 \times 10^{-3} \ m^{3}$ volume for unity power factor and 100\% load. Finally, the optimized design was simulated at 3 different power factor load and 4 different load rating, and results were presented for each.

\bibliographystyle{IEEEtran}
\bibliography{ref}

\end{document}